\documentclass{revtex4}
\usepackage{amsfonts}
\usepackage{amsmath}

\input{tcilatex}

\begin{document}

\preprint{}
\title{Solution for Static, Spherically Symmetric Lovelock Gravity Coupled
with Yang-Mills Hierarchy}
\author{S. Habib Mazharimousavi}
\email{habib.mazhari@emu.edu.tr}
\author{M. Halilsoy}
\email{mustafa.halilsoy@emu.edu.tr}
\affiliation{Department of Physics, Eastern Mediterranean University, G. Magusa, North
Cyprus, Mersin 10 - Turkey.}
\pacs{PACS number}

\begin{abstract}
The hierarchies of both Lovelock gravity and power-Yang-Mills field are
combined through gravity in a single theory. In static, spherically
symmetric ansatz exact particular integrals are obtained in all higher
dimensions. The advantage of such hierarchies is the possibility of choosing
coefficients, which are arbitrary otherwise, to cast solutions into
tractable forms. To our knowledge the solutions constitute the most general
spherically symmetric metrics that incorporate complexities both of Lovelock
and Yang-Mills hierarchies within the common context. A large portion of our
general class of solutions concern and addresses to black holes for which
specific examples are given. Thermodynamical behaviors of the system is
briefly discussed in particular dimensions.
\end{abstract}

\maketitle

\section{Introduction}

The hierarchy of Lovelock gravity consists of a sum ($\tsum\limits_{s=0}%
\alpha _{s}\tciLaplace _{s},$ $\alpha _{s}=$constant, $\tciLaplace
_{s}=s^{th}$ order Lagrangian) of geometrical terms representing higher
corrections in suitable combinations that do not give rise to equations
higher than second order \cite{1}. The higher order terms are reminiscent of
higher order Feynman diagrams in field theory but all at a classical level.
The zeroth order term ($s=0$) in the hierarchy is simply the cosmological
term while the first order ($s=1$) one corresponds to the familiar
Einstein-Hilbert (EH) Lagrangian. The second order ($s=2$) term gives the
Gauss-Bonnet (GB) gravity with the quadratic invariants. The third and
higher order Lovelock terms grow rather wildly, giving the impression that
it is impossible to keep the track analytically. Contrary to the
expectations, however, in particular geometries exact solutions are
available to all orders of the hierarchy. Not only the geometric terms but
with various sources, including power-Maxwell and power-Yang Mills (YM)
fields, exact solutions are available in static, spherical symmetric ansatz 
\cite{2,3}. By the power-Maxwell / YM, it is implied that the invariants in
the Lagrangian are raised to a power. The finely-tuned power has physical
implication as far as energy conditions are concerned \cite{3}. \ In
principle, $k$ can be chosen as an arbitrary $\left( \pm \right) $ rational
number, but such a freedom raises problems when the energy and causality
conditions are imposed. (Based on the energy conditions, $k$ must be at
least greater than $\frac{1}{2}$. Here in our study, since we aimed to
consider a discrete hierarchy, we restrict ourselves to the integer $k$
although this is not the only possible choice. In other words one may
consider a continues hierarchy with $\frac{1}{2}<k\in 
\mathbb{R}
$ which may be studied separately.) For this reason, to be on the safe side
we choose $k=\left( +\right) $ integer in this study. The topological
implication of such powers, if there is any at all, remains to be seen.

In this Letter, coupled with the Lovelock hierarchy we consider the YM
hierarchy (a different approach to YM hierarchy was first introduced by D.H.
Tchrakian in1985 \cite{4} and the concept was expanded later \cite{5}) of
the form $\sim \tsum\limits_{k}b_{k}\mathcal{F}^{k}$ where $b_{k}$ are
constant coefficients and $\mathcal{F}=$ the YM invariant $=F_{\mu \nu
}^{\left( a\right) }F^{\left( a\right) \mu \nu },$ with the internal index $%
a.$

It is interesting to note that for the YM invariant and dimension of
spacetime $d>5$, $\mathcal{F}\sim \frac{1}{r^{4}},$ irrespective of the
dimension. In the Maxwell case we recall that the invariant $\mathcal{F}%
_{M}\sim \frac{1}{r^{2\left( d-2\right) }},$ depends on the dimension as
well. (The reason that we excluded $d=5$ in the YM case is that it contains
a logarithmic term and violates the rule as aforesaid \cite{6}.) This
suggests, as a matter of fact, that we have a working YM hierarchy whereas
for the Maxwell case a similar hierarchy does not work with equal ease. In
obtaining an exact integral to the problem we make use of a Theorem proved
beforehand which is valid for a large class of energy-momenta \cite{7}.
Here, in particular we evaluate the integral for the general YM field
arising from the Wu-Yang ansatz \cite{8}. Let us add that it is this
particular ansatz which makes the YM hierarchy tractable in a diagonal
metric, simply by making the YM invariant mentioned above to have a fixed
power. It should be supplemented that the Wu-Yang ansatz in our choice works
only for the pure magnetic YM fields. Any other YM ansatz that can be
extended to higher dimensions analytically, even with a power (and
hierarchy), remains to be seen. The energy and causality conditions which
are employed in the Appendix determine the acceptable integers as a function
of dimensionality in our solution. These split naturally into two broad
classes labelled by 'even' and 'odd'. The intricate structure of our
solutions dashes hopes to determine horizons and thermodynamical functions
analytically. In principle, however, we obtain infinite class of solutions
pertaining to all dimensions that incorporate Lovelock and YM hierarchies in
the common metric. We choose particular parameters and dimensions to present
working examples of black hole solutions which elucidate our general class.
The $5-$dimensional black hole solution with an effective mass defined from
cosmological constant and YM charge is one such example. Chern-Simons (CS)
black hole solution in $d=11$ constitutes another example as an application
of our general class. From the definition of specific heat we show the
absence of thermodynamical phase transition for the CS black hole in $d=11$.

\section{$d$-dimensional Einstein-Lovelock Gravity with YM Hierarchy}

The $d-$dimensional action for Einstein-Lovelock-Yang-Mills-Hierarchies with
a cosmological constant $\Lambda$ is given by ($8\pi G=1$)

\begin{equation}
\begin{tabular}{l}
$I=\frac{1}{2}\int dx^{d}\sqrt{-g}\left( \tsum\limits_{s=0}^{\left[ \frac{d-1%
}{2}\right] }\alpha _{s}\tciLaplace _{s}-\tsum\limits_{k}^{q}b_{k}\mathcal{F}%
^{k}\right) $%
\end{tabular}%
\ \ \ \ \ ,
\end{equation}%
in which $\alpha _{0}=-\frac{\left( d-2\right) \left( d-1\right) }{3}\Lambda
,$ $\alpha _{1}=1,$ $\mathcal{F}$ is the YM invariant%
\begin{align}
\mathcal{F}& =\mathbf{\gamma }_{ab}(F_{\mu \nu }^{\left( a\right) }F^{\left(
b\right) \mu \nu }), \\
a,b& =1,2,...,\frac{\left( d-2\right) \left( d-1\right) }{2}\text{ \ and }%
\mathbf{\gamma }_{ab}=\delta _{ab}\text{, }  \notag
\end{align}%
The parameter $q$ ($1\leq k\leq q$) is an integer, $\alpha _{s}$ stand for
arbitrary constants, $\left[ \frac{d-1}{2}\right] $ represents the integer
part, and the Lovelock Lagrangian is 
\begin{equation}
\tciLaplace _{s}=2^{-n}\delta
_{c_{1}d_{1}...c_{n}d_{n}}^{a_{1}b_{1}...a_{n}b_{n}}R_{\ \
a_{1}b_{1}}^{c_{1}d_{1}}...R_{\ \ a_{s}b_{s}}^{c_{s}d_{s}},\text{ \ \ \ }%
s\geq 1.
\end{equation}%
Variation with respect to the gauge potentials $\mathbf{A}^{\left( a\right)
} $ yields the YM equations 
\begin{equation}
\tsum\limits_{k}b_{k}\left\{ \mathbf{d}\left( ^{\star }\mathbf{F}^{\left(
a\right) }\mathcal{F}^{k-1}\right) +\frac{1}{\sigma }C_{\left( b\right)
\left( c\right) }^{\left( a\right) }\mathcal{F}^{k-1}\mathbf{A}^{\left(
b\right) }\wedge ^{\star }\mathbf{F}^{\left( c\right) }\right\} =0.
\end{equation}%
where $^{\star }$ means duality, $C_{\left( b\right) \left( c\right)
}^{\left( a\right) }$ stands for the structure constants of $\frac{\left(
d-2\right) \left( d-1\right) }{2}-$ parameter Lie group $G,$ $\sigma $ is a
coupling constant and $\mathbf{A}^{\left( a\right) }$ are the $SO(d-1)$
gauge YM potentials. The determination of the components $C_{\left( b\right)
\left( c\right) }^{\left( a\right) }$ has been described elsewhere \cite{9}.
We note that the internal indices $\{a,b,c,...\}$ do not differ whether in
covariant or contravariant form. Variation of the action with respect to the
spacetime metric $g_{\mu \nu }$ yields the field equations

\begin{equation}
\tsum \limits_{s=0}^{\left[ \frac{d-1}{2}\right] }\alpha_{s}G_{\mu}^{\nu%
\left( s\right) }=T_{\mu}^{\nu},
\end{equation}
where 
\begin{equation}
T_{\ \nu}^{\mu}=-\frac{1}{2}\tsum \limits_{k}b_{k}\left( \delta_{\ \nu}^{\mu}%
\mathcal{F}^{k}-4k\mathbf{\gamma}_{ab}\left( F_{\nu\lambda}^{\left( a\right)
}F^{\left( b\right) \ \mu\lambda}\right) \mathcal{F}^{k-1}\right) .
\end{equation}
is the energy-momentum tensor representing the matter fields, and 
\begin{align}
G_{\mu}^{\nu\left( s\right) } & =\tsum \limits_{i=0}^{s}2^{-\left(
i+1\right) }\alpha_{i}\delta_{\mu c_{1}d_{1}...c_{i}d_{i}}^{\nu
a_{1}b_{1}...a_{i}b_{i}}R_{\ \ a_{1}b_{1}}^{c_{1}d_{1}}...R_{\ \
a_{i}b_{i}}^{c_{i}d_{i}},\text{ \ \ }s\geq1, \\
G_{\mu}^{\nu\left( 0\right) } & =\frac{\left( d-2\right) \left( d-1\right) }{%
6}\Lambda\delta_{\mu}^{\nu}\text{ , \ \ \ \ \ \ \ \ \ \ (}s=0\text{)}  \notag
\end{align}
Our metric ansatz for $d-$dimensions, is chosen as 
\begin{equation}
ds^{2}=-f\left( r\right) dt^{2}+\frac{dr^{2}}{f\left( r\right) }%
+r^{2}d\Omega_{\left( d-2\right) }^{2},
\end{equation}
in which $f\left( r\right) $ is our metric function. The choice of these
metrics can be traced back to the form of the stress-energy tensor (6),
which satisfies $T_{0}^{0}-T_{1}^{1}=0$ (see Eq. (12) below) and
consequently $G_{0}^{0}-G_{1}^{1}=0$, whose explicit form, on integration,
gives $\left\vert g_{00}g_{11}\right\vert =C=$constant. We need only to
choose the time scale at infinity to make this constant equal to unity.

Recently we have introduced and used the higher dimensional version of the
Wu-Yang \cite{8} ansatz in EYM theory of gravity \cite{8}. In this ansatz we
express the Yang-Mills magnetic gauge potential one-forms in the following
manner%
\begin{align}
\mathbf{A}^{(a)}& =\frac{Q}{r^{2}}C_{\left( i\right) \left( j\right)
}^{\left( a\right) }\ x^{i}dx^{j},\text{ \ \ }Q=\text{YM magnetic charge, \ }%
r^{2}=\overset{d-1}{\underset{i=1}{\sum }}x_{i}^{2}, \\
2& \leq j+1\leq i\leq d-1,\text{ \ and \ }1\leq a\leq \frac{\left(
d-2\right) \left( d-1\right) }{2},  \notag \\
x_{1}& =r\cos \theta _{d-3}\sin \theta _{d-4}...\sin \theta _{1},\text{ }%
x_{2}=r\sin \theta _{d-3}\sin \theta _{d-4}...\sin \theta _{1},  \notag \\
\text{ }x_{3}& =r\cos \theta _{d-4}\sin \theta _{d-5}...\sin \theta _{1},%
\text{ }x_{4}=r\sin \theta _{d-4}\sin \theta _{d-5}...\sin \theta _{1}, 
\notag \\
& ...  \notag \\
x_{d-1}& =r\cos \theta _{1}.  \notag
\end{align}%
One can easily show that these ansaetze satisfy the YM equations \cite{6,8}.
In consequence, the energy momentum tensor (6), with 
\begin{align}
\mathcal{F}& =\frac{\left( d-2\right) \left( d-1\right) Q^{2}}{r^{4}}, \\
\mathbf{Tr}\left( F_{\theta _{i}\lambda }^{\left( a\right) }F^{\left(
a\right) \ \theta _{i}\lambda }\right) & =\frac{\left( d-3\right) Q^{2}}{%
r^{4}}=\frac{1}{d-2}\mathcal{F}
\end{align}%
takes the compact form%
\begin{equation}
T_{\text{ }\nu }^{\mu }=-\frac{1}{2}\tsum\limits_{k}b_{k}\mathcal{F}^{k}%
\text{diag}\left[ 1,1,\xi ,\xi ,..,\xi \right] ,\text{ \ with \ }\xi =\left(
1-\frac{4k}{d-2}\right) .
\end{equation}

\subsection{Energy conditions and the solutions}

Upon choosing the energy momentum tensor, it is necessary to look at the
energy conditions. This is important, because the upper and lower limits of $%
k$ will come to light by imposing the energy and causality conditions all
satisfied. In a straightforward calculation (see the Appendix) one can show
that WEC, SEC, DEC and CC are all satisfied if and only if $\frac{d-1}{4}%
\leq k<\frac{d-1}{2}.$ Therefore we should modify our summation symbol
accordingly as%
\begin{equation}
\tsum\limits_{k}b_{k}\rightarrow \tsum\limits_{k=-[-\frac{d-1}{4}]}^{-[-%
\frac{d-1}{2}]-1}b_{k}.
\end{equation}%
Here one should notice that in 4 and 5 dimensions only $b_{1}$ is available
and for 6 and 7 dimensions $b_{2}$ is nonzero. Of course, for $d-$dimensions
we have $[-\frac{d-1}{4}]-[-\frac{d-1}{2}]$ terms included. Our static,
spherically symmetric metric is given by (8), whose metric function can be
reexpressed, for convenience in the form 
\begin{equation}
f\left( r\right) =1-r^{2}H\left( r\right) ,
\end{equation}%
and from the $tt$ component of\ (5) and (12) we obtain \cite{7}%
\begin{equation}
\tsum\limits_{s=0}^{\left[ \frac{d-1}{2}\right] }\tilde{\alpha}_{s}H^{s}=%
\frac{4m}{\left( d-2\right) r^{d-1}}-\frac{2}{\left( d-2\right) r^{d-1}}%
\tint r^{d-2}T_{t}^{t}dr.
\end{equation}%
Here $m$ is an integration constant related to the ADM mass of the black
hole, $\tilde{\alpha}_{0}=-\frac{\Lambda }{3},$ $\tilde{\alpha}_{1}=1,$ and 
\begin{equation}
\tilde{\alpha}_{s}=\overset{2s}{\underset{i=3}{\Pi }}\left( d-i\right)
\alpha _{s}\text{, }s>1.
\end{equation}%
Now, we use $T_{t}^{t}$ given in (12) to get 
\begin{equation}
\tsum\limits_{s=0}^{[\frac{d-1}{2}]}\tilde{\alpha}_{s}H^{s}=\frac{4m}{\left(
d-2\right) r^{d-1}}+\tsum\limits_{k=-[-\frac{d-1}{4}]}^{-[-\frac{d-1}{2}]-1}%
\frac{b_{k}\tilde{Q}_{k}^{2}}{\left( d-2\right) }\left\{ 
\begin{array}{lr}
\frac{1}{\left( d-1-4k\right) r^{4k}}, & k\neq \frac{d-1}{4} \\ 
\frac{\ln r}{r^{d-1}}, & k=\frac{d-1}{4}%
\end{array}%
\right. =\Psi .
\end{equation}%
where $\tilde{Q}_{k}^{2}=\left( \left( d-2\right) \left( d-1\right)
Q^{2}\right) ^{k}$and $\Psi $ abbreviates the indicated series. Here we
comment that at $r\rightarrow \infty ,$ one gets%
\begin{equation}
\lim_{r\rightarrow \infty }\Psi =\left. \frac{b_{k}\tilde{Q}_{k}^{2}}{\left(
d-2\right) }\left\{ 
\begin{array}{lr}
\frac{1}{\left( d-1-4k\right) r^{4k}}, & d\neq 5,9,13,17,... \\ 
\frac{\ln r}{r^{d-1}}, & d=5,9,13,17,...%
\end{array}%
\right. \right\vert _{k=-[-\frac{d-1}{4}]}.
\end{equation}%
Let's introduce new parameters as 
\begin{equation}
\tilde{\alpha}_{s}=\frac{\bar{\alpha}_{s}}{\bar{\alpha}_{1}},\text{ for }%
s\geq 2\text{ and }-\frac{\Lambda }{3}=\frac{\bar{\alpha}_{0}}{\bar{\alpha}%
_{1}},
\end{equation}%
which lead to 
\begin{equation}
\begin{tabular}{l}
$\tsum\limits_{s=0}^{[\frac{d-1}{2}]}\bar{\alpha}_{s}H^{s}=\bar{\alpha}%
_{1}\Psi $%
\end{tabular}%
\end{equation}%
and choose a specific set of \cite{10} $\bar{\alpha}_{s}$ such that 
\begin{equation}
\bar{\alpha}_{s}=\left( \pm 1\right) ^{s+1}\binom{\left[ \frac{d-1}{2}\right]
}{s}\ell ^{2s-d}
\end{equation}%
where $-\frac{\Lambda }{3}=\frac{\bar{\alpha}_{0}}{\bar{\alpha}_{1}}=\pm 
\frac{\ell ^{-2}}{\left[ \frac{d-1}{2}\right] }.$ Following the latter
expression, Eq. (20) gives 
\begin{equation}
\begin{tabular}{l}
$\left( 1\pm \ell ^{2}H\right) ^{\left[ \frac{d-1}{2}\right] }=\pm \ell ^{d}%
\bar{\alpha}_{1}\Psi $%
\end{tabular}%
\end{equation}%
and consequently%
\begin{equation}
f^{\left( \pm \right) }\left( r\right) =1\pm \frac{r^{2}}{\ell ^{2}}\mp 
\frac{r^{2}}{\ell ^{2}}\sigma \left( \pm \left[ \frac{d-1}{2}\right] \ell
^{2}\Psi \right) ^{1/\left[ \frac{d-1}{2}\right] },
\end{equation}%
in which 
\begin{equation}
\sigma =\left\{ 
\begin{array}{lr}
\pm 1, & [\frac{d-1}{2}]=even \\ 
1, & [\frac{d-1}{2}]=odd%
\end{array}%
\right. .
\end{equation}%
After this general solution we label the solutions for even and odd
dimensions separately. To do so, we put $\left[ \frac{d-1}{2}\right] =\frac{%
d-1}{2}$ for odd dimensions and $\left[ \frac{d-1}{2}\right] =\frac{d-2}{2}$
for even dimensions into (23) to obtain the splitting 
\begin{equation}
f_{even}^{\left( \pm \right) }\left( r\right) =1\pm \frac{r^{2}}{\ell ^{2}}%
\mp \sigma \left[ \pm \frac{d-2}{2\ell ^{d-4}}\left( \frac{4m}{\left(
d-2\right) r}+\tsum\limits_{k=-[-\frac{d-1}{4}]}^{-[-\frac{d-1}{2}]-1}\frac{%
b_{k}\tilde{Q}_{k}^{2}}{\left( d-2\right) }\left\{ 
\begin{array}{lr}
\frac{1}{\left( d-1-4k\right) r^{4k-d+2}}, & k\neq \frac{d-1}{4} \\ 
\frac{\ln r}{r}, & k=\frac{d-1}{4}%
\end{array}%
\right. \right) \right] ^{\frac{2}{d-2}},
\end{equation}%
and 
\begin{equation}
f_{odd}^{\left( \pm \right) }\left( r\right) =1\pm \frac{r^{2}}{\ell ^{2}}%
\mp \sigma \left[ \pm \frac{d-1}{2\ell ^{d-3}}\left( \frac{4m}{\left(
d-2\right) }+\tsum\limits_{k=-[-\frac{d-1}{4}]}^{-[-\frac{d-1}{2}]-1}\frac{%
b_{k}\tilde{Q}_{k}^{2}}{\left( d-2\right) }\left\{ 
\begin{array}{lr}
\frac{1}{\left( d-1-4k\right) r^{4k-d+1}}, & k\neq \frac{d-1}{4} \\ 
\ln r, & k=\frac{d-1}{4}%
\end{array}%
\right. \right) \right] ^{\frac{2}{d-1}}.
\end{equation}%
From $f_{even}^{\left( \pm \right) }\left( r\right) ,$ for instance the
Einstein-de Sitter limit can readily be seen for $d=4$ and $Q_{k}=0.$ It is
remarkable to observe that by setting $b_{k}=0$ we obtain 
\begin{equation}
\left. f_{even}^{\left( \pm \right) }\left( r\right) \right\vert
_{b_{k}=0}=1\pm \frac{r^{2}}{\ell ^{2}}\mp \sigma \left[ \pm \frac{1}{\ell
^{d-4}}\frac{2m}{r}\right] ^{\frac{2}{d-2}}
\end{equation}%
and%
\begin{equation}
\left. f_{odd}^{\left( \pm \right) }\left( r\right) \right\vert
_{b_{k}=0}=1\pm \frac{r^{2}}{\ell ^{2}}\mp \sigma \left[ \pm \frac{1}{\ell
^{d-3}}\left( \frac{2\left( d-1\right) m}{\left( d-2\right) }\right) \right]
^{\frac{2}{d-1}}
\end{equation}%
which by choosing the positive branches and redefinition of the free
parameters we get the results reported in \cite{11}. Therefore we use only
the positive branches for our further study. Here we investigate the
possible horizon of the above black hole solutions.

\subsubsection{Even dimensions}

To find the horizon(s) of the solution given in Eq. (25) we set%
\begin{equation}
f_{even}^{\left( +\right) }\left( r_{h}\right) =0,
\end{equation}%
which admits the relation between the black hole's parameters. Finding
horizon(s) in a closed form is not possible, therefore we choose a specific
dimension, namely $d=8$ for going further. In this setting the latter
equation reads 
\begin{equation}
1+\frac{r_{h}^{2}}{\ell ^{2}}-\left[ \frac{1}{\ell ^{4}}\left( \frac{2m}{%
r_{h}}-\frac{b_{2}\tilde{Q}_{2}^{2}}{2}\frac{1}{r_{h}^{2}}-\frac{b_{3}\tilde{%
Q}_{3}^{2}}{10}\frac{1}{r_{h}^{6}}\right) \right] ^{\frac{1}{3}}=0
\end{equation}%
Fig. 1 displays $\rho =\frac{r_{h}}{\ell }$ in terms of $\mu =$ $\frac{b_{2}%
\tilde{Q}_{2}^{2}}{2\ell ^{6}},\nu =\frac{b_{3}\tilde{Q}_{3}^{2}}{10\ell
^{10}}$ for $\frac{m}{\ell ^{5}}=1$. Depending on $\mu $ and $\nu $, two
horizons or no horizon cases are the basic information that Fig. 1 reveals.
Changing $m$ does not effect the general schema of the figure. From the
metric one observes that $r=0$ is a singularity hidden by horizon(s) and the
Ricci scalar, once $r\rightarrow 0$ behaves as%
\begin{equation}
\lim_{r\rightarrow 0}R\rightarrow \frac{-12\nu ^{1/3}}{\ell ^{2/3}r^{4}}%
\rightarrow \pm \infty .
\end{equation}%
.

\subsubsection{ Odd dimensions}

Again, in this part, we set the metric function (26) to zero, i.e.,%
\begin{equation}
f_{odd}^{\left( +\right) }\left( r_{h}\right) =0
\end{equation}%
which after we choose a specific odd dimension, namely $d=9$ it reads%
\begin{equation}
1+\frac{r_{h}^{2}}{\ell ^{2}}-\sigma \left[ \frac{1}{\ell ^{6}}\left( \frac{%
16m}{7}+\frac{4b_{2}\tilde{Q}_{2}^{2}}{7}\ln r_{h}-\frac{b_{3}\tilde{Q}%
_{3}^{2}}{7}\frac{1}{r_{h}^{4}}\right) \right] ^{\frac{1}{4}}=0.
\end{equation}%
Unlike the previous example, here $\sigma =\pm 1$ but for $\sigma =-1$
definitely there is no horizon and our solution collapses to a cosmological
object which is not of interest. For $\sigma =+1$ the solution admits black
hole with horizon(s). In Fig. 2 we plot $\rho =\frac{r_{h}}{\ell }$ in terms
of $\mu =$ $\frac{4b_{2}\tilde{Q}_{2}^{2}}{7\ell ^{6}},\nu =\frac{b_{3}%
\tilde{Q}_{3}^{2}}{7\ell ^{10}}$ and for $\frac{16m}{7\ell ^{6}}=\mu \frac{%
\ln \ell ^{2}}{\ell ^{6}}+1$. We observe that Fig.2 shares much of the
features with Fig. 1. One should notice that in this case we have the
condition 
\begin{equation}
\frac{1}{\ell ^{6}}\left( \frac{16m}{7}+\frac{4b_{2}\tilde{Q}_{2}^{2}}{7}\ln
r_{h}-\frac{b_{3}\tilde{Q}_{3}^{2}}{7}\frac{1}{r_{h}^{4}}\right) \geq 0.
\end{equation}%
Although these two examples are given in specific dimensions, they show how
the procedure goes on and definitely in higher dimensions having more terms
in the hierarchy makes the analysis much more complicated. Let us add that
the Ricci scalar in this case diverges as $r^{-3\text{ }}$ which shows that $%
r=0$ is a singular point hidden behind the horizon(s).

\subsection{A very specific case}

Now let us relax the energy conditions except the WEC which allows us to
choose $k=0,1,...,[\frac{d-1}{2}].$ For the case of $k\neq \frac{d-1}{4}$
one finds from (17) that%
\begin{equation}
\tsum\limits_{s=0}^{[\frac{d-1}{2}]}\bar{\alpha}_{s}H^{s}=\bar{\alpha}%
_{1}\left( \frac{4m}{\left( d-2\right) r^{d-1}}+\tsum\limits_{k=0}^{[\frac{%
d-1}{2}]}b_{k}\frac{\tilde{Q}_{k}^{2}}{\left( d-2\right) \left(
d-1-4k\right) r^{4k}}\right)
\end{equation}%
which after setting $m=0$ and $\frac{b_{k}\tilde{Q}_{k}^{2}}{\left(
d-2\right) \left( d-1-4k\right) }=\beta _{k}=\left( \pm 1\right) ^{k+1}%
\binom{[\frac{d-1}{2}]}{k}\lambda ^{2k-d}$ this admits 
\begin{equation}
\tsum\limits_{s=0}^{[\frac{d-1}{2}]}\bar{\alpha}_{s}H^{s}=\bar{\alpha}%
_{1}\tsum\limits_{k=0}^{[\frac{d-1}{2}]}\beta _{k}\left( \frac{1}{r^{4}}%
\right) ^{k}.
\end{equation}%
This yields 
\begin{equation}
\tsum\limits_{s=0}^{[\frac{d-1}{2}]}\left( \pm 1\right) ^{s}\binom{[\frac{d-1%
}{2}]}{s}\ell ^{2s-d}H^{s}=\tsum\limits_{k=0}^{[\frac{d-1}{2}]}\left( \pm
1\right) ^{k}\binom{[\frac{d-1}{2}]}{k}\lambda ^{2k-d}\left( \frac{1}{r^{4}}%
\right) ^{k}
\end{equation}%
or%
\begin{equation}
\ell ^{-d}\left( 1\pm \ell ^{2}H\right) ^{[\frac{d-1}{2}]}=\bar{\alpha}%
_{1}\lambda ^{-d}\left( 1\pm \frac{\lambda ^{2}}{r^{4}}\right) ^{[\frac{d-1}{%
2}]}
\end{equation}%
which, after adjusting $\bar{\alpha}_{1}\lambda ^{-d}=\ell ^{-d}$ one finds%
\begin{equation}
\left( 1\pm \ell ^{2}H\right) ^{[\frac{d-1}{2}]}=\left( 1\pm \frac{\lambda
^{2}}{r^{4}}\right) ^{[\frac{d-1}{2}]}
\end{equation}%
and depending on the dimensionality we have%
\begin{equation}
1\pm \ell ^{2}H=\sigma \left( 1\pm \frac{\lambda ^{2}}{r^{4}}\right) .
\end{equation}%
This leads to%
\begin{equation}
H=\mp \frac{1}{\ell ^{2}}\pm \frac{\sigma }{\ell ^{2}}\left( 1\pm \frac{%
\lambda ^{2}}{r^{4}}\right) ,
\end{equation}%
and consequently%
\begin{equation}
f\left( r\right) =1\pm \frac{r^{2}}{\ell ^{2}}\mp \frac{\sigma }{\ell ^{2}}%
\left( r^{2}\pm \frac{\lambda ^{2}}{r^{2}}\right) =\left\{ 
\begin{tabular}{ll}
$1-\frac{\lambda ^{2}}{\ell ^{2}}\frac{1}{r^{2}},$ & $d=7,8,11,12,15,16,...$
\\ 
$1-\frac{\lambda ^{2}}{\ell ^{2}}\frac{1}{r^{2}}$ and $1\pm \frac{2}{\ell
^{2}}r^{2}+\frac{\lambda ^{2}}{\ell ^{2}}\frac{1}{r^{2}},$ & $%
d=5,6,9,10,13,14,...$%
\end{tabular}%
\ \ \ \ \ \ \ \ \ \right. .
\end{equation}%
It is remarkable to observe that the latter solution is nothing but the
Schwarzschild black hole-like solution in $5-$dimensions if we consider $%
\frac{\lambda ^{2}}{\ell ^{2}}$ as the effective mass of the black hole.
Note that the mass term of the black hole, $m$ was chosen to be zero. Also,
for the other set of solutions i.e.%
\begin{equation}
f\left( r\right) =1\pm \frac{2}{\ell ^{2}}r^{2}+\frac{\lambda ^{2}}{\ell ^{2}%
}\frac{1}{r^{2}},
\end{equation}%
one may call it anti-Schwarzschild black hole with a positive or negative
cosmological constant. To get a better idea about this solution we rewrite
it in terms of $m_{eff}=\frac{\lambda ^{2}}{\ell ^{2}}$ and $\Lambda
_{eff}=\pm \frac{2}{\ell ^{2}},$ so that%
\begin{equation*}
f\left( r\right) =1+\Lambda _{eff}r^{2}+\frac{m_{eff}}{r^{2}}.
\end{equation*}%
Let us remind, from the above identifications, that $m_{eff}$ depends on
both $\ell $ and $Q_{k}.$ It is clear that with positive sign there is no
horizon and therefore it is a cosmological object which has a naked
singularity at the origin. The negative branch has a cosmological horizon at%
\begin{equation}
r_{h}=\left( \frac{\ell ^{2}+\sqrt{\ell ^{2}+8\lambda ^{2}}}{4}\right)
^{1/2}.
\end{equation}
\ 

\subsection{Example of Chern-Simons (CS) gravity in $11-$dimensions}

As one may notice, setting the $[\frac{d-1}{2}]$ Lovelock parameters
according to (21), in odd dimensions it becomes isometric with the CS theory
of gravity \cite{10,11,12}. Therefore Eq. (26) gives a black hole solution
in CS theory, and in this section we shall go through some of the physical
properties of this type BHs in $11-$dimensions as an example.

\subsubsection{For $k\neq\frac{d-1}{4}$ with positive branch}

The solution, after choosing $\ell ^{-2}=-[\frac{d-1}{2}]\frac{\Lambda }{3}%
=1 $ and rewriting the integration constants, in $11-$dimensions, reads%
\begin{equation}
f_{odd}\left( r\right) =1+r^{2}-\left( 1+M-\frac{\mu }{r^{2}}-\frac{\nu }{%
r^{6}}\ \ \right) ^{\frac{1}{5}}
\end{equation}%
in which $\mu =202500b_{3}Q^{6}$ , $\nu =6075000b_{4}Q^{8}$ and $M=\frac{20}{%
9}m$. We remark that although the constants $\mu $ and $\nu $ are
multipole-like coefficients depending on powers of the YM charge $Q$ and
cosmological constant, which is scaled to unity, their exact interpretation
can be understood upon expansion of the power. From the energy conditions
(see the Appendix) we show that $b_{k}\geq 0;$ this implies restriction on
the mass parameter $M$ so that the parenthesis in (45) is positive. The
Hawking temperature and the mass of the black hole are given by 
\begin{equation}
T_{H}=\frac{1}{4\pi }f^{\prime }\left( r_{h}\right) =\frac{r_{h}}{2\pi }-%
\frac{\mu }{10\pi r_{h}^{3}\left( 1+r_{h}^{2}\right) ^{4}}-\frac{3\nu }{%
10\pi r_{h}^{7}\left( 1+r_{h}^{2}\right) ^{4}},
\end{equation}%
and%
\begin{equation}
M=\frac{\mu }{r_{h}^{2}}+\frac{\nu }{r_{h}^{6}}+\left( r_{h}^{2}+1\right)
^{5}-1,
\end{equation}%
respectively. The specific heat \cite{12} 
\begin{equation}
C_{Q}=\left( \frac{\partial M}{\partial T_{H}}\right) _{Q},
\end{equation}%
reads as%
\begin{equation}
C_{Q}=-20\frac{\pi r_{h}\left( 1+r_{h}^{2}\right) ^{5}\left[ r_{h}^{4}\left(
\mu -5r_{h}^{4}\left( 1+r_{h}^{2}\right) ^{4}\right) +3\nu \right] }{%
r_{h}^{2}\left\{ 3\mu r_{h}^{2}+45\nu +r_{h}^{4}\left[ 11\mu
+5r_{h}^{2}\left( 1+r_{h}^{2}\right) ^{5}\right] \right\} +21\nu }.
\end{equation}%
We observe that absence of root(s) of the denominator implies that the CSBH
does not experience phase changes.

\subsubsection{For $k\neq\frac{d-1}{4}$ with negative branch}

By a similar setting as in the previous subsection, after choosing the
negative branch of the solution one gets%
\begin{equation}
f_{odd}\left( r\right) =1-r^{2}+\left( 1+M+\frac{\mu }{r^{2}}+\frac{\nu }{%
r^{6}}\ \ \right) ^{\frac{1}{5}}.
\end{equation}%
We note that the integration constant $M$ and the parameters $\mu ,\nu $
have the same values as in Eq. (45). In this branch it is readily seen that
there is no restriction on $M$, since the expression in the parenthesis is
always positive. In this case also we use the same definitions to find 
\begin{equation}
T_{H}=\frac{1}{4\pi }f^{\prime }\left( r_{h}\right) =-\frac{r_{h}}{2\pi }-%
\frac{\mu }{10\pi r_{h}^{3}\left( 1+r_{h}^{2}\right) ^{4}}-\frac{3\nu }{%
10\pi r_{h}^{7}\left( 1+r_{h}^{2}\right) ^{4}},
\end{equation}%
and%
\begin{equation}
M=\frac{\mu }{r_{h}^{2}}+\frac{\nu }{r_{h}^{6}}+\left( r_{h}^{2}-1\right)
^{5}-1
\end{equation}%
\begin{equation}
C_{Q}=20\frac{\pi r_{h}\left( r_{h}^{2}-1\right) ^{5}\left[ r_{h}^{4}\left(
\mu -5r_{h}^{4}\left( r_{h}^{2}-1\right) ^{4}\right) +3\nu \right] }{%
r_{h}^{2}\left\{ -3\mu r_{h}^{2}+45\nu +r_{h}^{4}\left[ 11\mu
+5r_{h}^{2}\left( r_{h}^{2}-1\right) ^{5}\right] \right\} -21\nu }.
\end{equation}%
The zeros of the denominator implies possible phase changes in the CSBH,
however, the fact that $T_{H}<0$ makes this particular case questionable.

\section{Conclusion}

With the exception of highly symmetric cases finding general integrals to
Einstein's field equations in general relativity remained ever challenging.
Add to that the most general Lovelock gravity and YM hierarchies, doubtless
makes it further challenging. By resorting to a previously known theorem in
generating solutions and simplicity of power-YM theory / hierarchy aided in
obtaining such particular integrals. The reported static, spherically
symmetric metrics are valid in all higher dimensions and occurrence of
polynomials with rational powers in closed form seems to be their
characteristic feature. A particular example refers to the $11-$dimensional
Chern-Simons (CS) gravity in which the intricacy of the metric function is
clearly seen. Determination of zeros of such a function remains a
mathematical challenge. For particular dimensions, i.e. d=8,9, we plot in
Fig.s 1 and 2 explicit formation of horizons. From the thermodynamical
analysis we evaluate the relevant quantities and investigate the possibility
of phase transitions in this model. One particular example that yields $%
T_{H}<0$, must be discarded as non-physical. The causality and energy
conditions discussed in Appendix guide us to fix the acceptable dimensions
for each particular case.

\begin{acknowledgments}
We are indebted to the anonymous referee for drawing our attention to some
erroneous statements in the original version of the paper.
\end{acknowledgments}

\textbf{APPENDIX: Energy Conditions}

When a matter field couples to any system, energy conditions must be
satisfied for physically acceptable solutions. We follow the steps as given
in \cite{9}.

\subsection{Weak Energy Condition (WEC)}

\begin{equation}
T_{\text{ }\nu}^{\mu}=-\frac{1}{2}\tsum \limits_{k=1}^{q}b_{k}\mathcal{F}^{k}%
\text{diag}\left[ 1,1,\xi,\xi,..,\xi\right] ,\text{ \ and \ }\xi=\left( 1-%
\frac{4k}{d-2}\right)
\end{equation}

The WEC states that,

\begin{equation}
\rho\geq0\text{ \ \ \ \ \ \ \ \ \ \ \ and \ \ \ \ \ \ \ \ }\rho+p_{i}\geq0%
\text{ \ \ \ \ \ }(i=1,2,...d-1)  \tag{A1}
\end{equation}
in which $\rho$ is the energy density and $p_{i}$ are the principal
pressures given by

\begin{equation}
\rho=-T_{t}^{t}=-T_{r}^{r}=\frac{1}{2}\tsum \limits_{k}b_{k}\mathcal{F}^{k},%
\text{ \ \ \ \ \ \ \ \ }p_{i}=T_{i}^{i}\text{ \ \ (no sum)}  \tag{A2}
\end{equation}
The WEC imposes the following conditions on the constant parameters $b_{k}$
and $k$;

\begin{equation}
0\leq b_{k}\text{ \ \ \ \ \ and \ \ \ \ \ }0\leq k,  \tag{A3}
\end{equation}

\subsection{Strong Energy Condition (SEC)}

This condition states that;

\begin{equation}
\rho+\dsum \limits_{i=1}^{d-1}p_{i}\geq0\text{ \ \ \ \ \ \ and \ \ \ \ \ \ }%
\rho+p_{i}\geq0.  \tag{A4}
\end{equation}
This condition together with the WEC constrain the parameters as,

\begin{equation}
0\leq b_{k}\text{ \ \ \ \ \ and \ \ \ \ }\frac{\text{\ }d-2}{4}\leq k. 
\tag{A5}
\end{equation}

\subsection{Dominant Energy Condition (DEC)}

In accordance with DEC, the effective pressure $p_{eff}$ should not be
negative i.e. $p_{eff}\geq0$ where

\begin{equation}
p_{eff}=\frac{1}{d-1}\dsum \limits_{i=1}^{d-1}T_{i}^{i}.  \tag{A6}
\end{equation}
One can show that DEC, together with SEC and WEC impose the following
conditions on the parameters 
\begin{equation}
0\leq b_{k}\text{ \ \ \ \ \ and \ \ \ \ \ }\frac{d-1}{4}\leq k.  \tag{A7}
\end{equation}

\subsection{Causality Condition (CC)}

In addition to the energy conditions one can impose the causality condition
(CC)

\begin{equation}
0\leq\frac{p_{eff}}{\rho}<1,  \tag{A8}
\end{equation}
which implies

\begin{equation}
0\leq b_{k}\text{ \ \ \ \ \ and \ \ \ \ }\frac{\text{\ }d-1}{4}\leq k\leq 
\frac{\text{\ }d-1}{2}.  \tag{A9}
\end{equation}

\textbf{Figure Captions:}

Fig. 1: The 3-dimensional parametric plot for Eq. (30), i.e. $f(r_{h})=0$.
Plotting of $\rho =\frac{r_{h}}{\ell }$ versus $\mu $ and $\nu $ referring
to even ($d=8$) dimensions is given. The occurrence of two horizons / no
horizon is clearly visible. The fact that we abide by $\mu >0$ and $\nu >0$,
originates from the energy conditions which dictates $b_{k}\geq 0.$ It can
easily be seen that $\mu $ plays little role in comparison with $\nu .$

Fig. 2: Plotting of $\rho =\frac{r_{h}}{\ell }$ versus $\mu $ and $\nu $
from Eq. (33). We have again dominantly two or no horizon cases. For small $%
\nu $ values we have rare formation of single horizon. The effect of $\nu $
dominates over $\mu $ also here for odd ($d=9$) dimensions.

\bigskip

\end{document}